\documentclass[11pt]{article}
\def \beq  {\begin{equation}}
\def \eeq  {\end{equation}}
\def \beqar {\begin{eqnarray}}
\def \eeqar {\end{eqnarray}}

\def\sqr#1#2{{\vcenter{\vbox{\hrule height.#2pt
\hbox{\vrule width.#2pt height#1pt \kern#1pt
\vrule width.#2pt}\hrule height.#2pt}}}}

\def\la {{\langle}}
\def\ra {{\rangle}}

\def\vf {{\varphi}}

\def\Tr {{\rm Tr}}

\def \dotA {{\dot A}}
\def \dotB {{\dot B}}

\def\bD {\bar{D}}
\def\bA {\bar{A}}

\def\bu {\bar{u}}

\def\del {\partial}
\def\bdel{\bar{\partial}}

\def\vf {{\varphi}}

\begin{document}

\begin{titlepage}
\null\vspace{-62pt} \pagestyle{empty}
\begin{center}
\rightline{CCNY-HEP-05/1}
\rightline{January 2005}
\vspace{1truein} {\Large\bf
A Note on MHV Amplitudes for Gravitons}
\vskip .2in\noindent

\vspace{.6in}
V. P.~NAIR
\vskip .1in
{\it Physics Department\\
City College of the CUNY\\
New York, NY 10031}\\
E-mail: vpn@sci.ccny.cuny.edu\\
\vskip .1in

\vspace{.4in}
\centerline{\large\bf Abstract}
\end{center}
We show how the maximally helicity violating (MHV) 
scattering amplitudes for gravitons can be related to
current correlators and vertex operators
in twistor space. This is similar to what happens in 
Yang-Mills theory and raises the possibility of a 
direct twistor-string-like construction for
${\cal N}=8$ supergravity.

\end{titlepage}
\pagestyle{plain} \setcounter{page}{2}
Recently, there has been a lot of progress in the calculation and 
understanding of multigluon scattering amplitudes in the
${\cal N}=4$ super Yang-Mills theory based on ideas of twistor string theory\cite{witten}.
The simplest set of such amplitudes are the 
so-called maximally helicity violating (MHV) ones, calculated by standard
field theory techniques a 
number of years ago
\cite{pt}.
The mathematical expressions for these amplitudes turned out to be very
simple and concise, although the intermediate steps of the calculation were
algebraically very involved.
The MHV amplitudes could also be interpreted in terms of
the
current correlators of a Wess-Zumino-Witten (WZW)
theory (or using free fermions); this had a natural
interpretation in supertwistor space \cite{nair}.
Witten showed the deep connection of these results
to twistor string theory, where the target manifold is 
the supertwistor
space ${\bf CP}^{3|4}$, which is a Calabi-Yau space.
A topological
version of such a string theory, the 
topological $B$-model, is what is relevant for the
Yang-Mills amplitudes.
The MHV amplitude is the restriction
of a holomorphic function in ${\bf CP}^{3|4}$ to
a complex line.
This complex line can be interpreted as a $D$-instanton in the
string theory.
The correlators of the $B$-model on this line become WZW correlators,
reproducing the MHV amplitudes.
Witten also argued that
the non-MHV amplitudes can be obtained
by considering
algebraic curves of higher degree in ${\bf CP}^{3|4}$.
This has been verified in a number of specific cases \cite{list1}. 
Later, it was realized that there could be even more dramatic simplifications
\cite{witten3}.
By treating the MHV amplitudes as the basic vertices,
with a suitable
off-shell continuation, and joining them
together via propagators, one could obtain all Yang-Mills amplitudes.
This procedure could be thought of as a limiting case of the 
higher degree curves, where the curves degenerate into intersecting 
(complex) lines.
The calculation of many amplitudes using this CSW procedure has been carried out
and agrees with known Yang-Mills amplitudes, for those cases where
a comparison with other techniques is possible \cite{list2}. 
It is worth emphasizing that some loop calculations have also been done using these
methods \cite{spence}. Twistor string theory is thus very remarkable,
giving scattering amplitudes of the gauge theory by a simple
set of rules in the twistor language.

A different type of twistor string theory which leads to the
same amplitudes
has been proposed by Berkovits \cite{berk}.
Rather than a $D$-instanton calculation, the amplitudes are now given 
by  the standard string perturbation expansion.

Given these remarkable developments, which are all related to the Yang-Mills theory, 
a natural question would be whether there is a parallel to this
for graviton scattering. Is there a twistor description for
perturbative graviton scattering? One of the motivations for twistor string theory 
was to seek a gauge-gravity duality at weak coupling \cite{witten}, and from that point of view, 
graviton amplitudes are somewhat different, being part of the
dual description. Nevertheless, it is interesting to see how far
a parallel can be developed; this may 
give new insights into various relationships between
string theory and gauge theory. In this paper, we shall consider the 
tree level MHV amplitudes for
gravitons.
These were calculated many years ago by Berends, Giele and Kuijf (BGK)
using the field theory limit
of the Kawai-Lewellyn-Tye (KLT) relations between closed string amplitudes 
and open string amplitudes \cite{berends}.
The BGK formulae and the KLT relations were also used by
Bern {\it et al} to analyze loop amplitudes in gravity \cite{bern1}.
The formulae involve many nonholomorphic factors in terms of the spinor momenta
and so their twistor space description will be necessarily more complicated than the Yang-Mills case. One could represent the nonholomorphic terms as derivatives
with respect to the twistor variables\cite{witten}.
The result is then not so elegant as the gauge theory result, and so, we shall
try a different approach.
We will carry out some simplification of the MHV amplitudes and show that
the nonholomorphic factors can be interpreted as the analog of the 
Chan-Paton factors of
the open string/gauge theory case. ${\cal N}=8$ supergravity is the natural
theory for carrying out the simplifications. Our results are suggestive of a 
twistor-string-like construction which leads to ${\cal N}=8$ supergravity.
It is interesting to note at this point that gravitons arise from open string amplitudes
in Berkovits' string theory, so that it may be natural to incorporate Chan-Paton factors
in some extension of this theory.
Twistorial interpretation of gravity amplitudes in terms of localization
on curves has been analyzed to some extent in \cite{giombi}.
The possibility of a current which incorporates gravitons has been
considered in \cite{seliv}.

We shall begin with a brief recapitulation of the MHV amplitudes for Yang-Mills fields
in terms of current correlators or expectation values of vertex operators \cite{nair}.
Since the particles of interest
are massless, they have four-momenta $p_\mu$ which are null, i.e.,
$p^2 =0$. As is well known, such a momentum can be written in terms of
spinors as
$p^A_{~\dotA } = (\sigma^\mu)^A_{~\dotA }
~p_\mu= u^A \bu_{\dotA} $.
For real four-momenta $\bu_{\dotA} =(u^A)^*$. 
In much of the recent work
using twistor string theory, this condition is relaxed since it is 
more convenient to consider complex momenta, or other reality conditions corresponding to
different choices of spacetime signature.

There is a natural action of the Lorentz group $SL(2,{\bf C})$
on the dotted and undotted indices, given by
\beq
u^A \rightarrow u'^A =
(g u)^A, \hskip .3in
\bu_{\dotA} \rightarrow \bu'_{\dotA}
= (g^{*}\bu)_{\dotA}
\label{2}
\eeq
where $g \in SL(2, {\bf C})$.
The Lorentz-invariant scalar products of the spinor-momenta
are given by $\la 1 2\ra= \epsilon_{AB} u^A_1
u^B_2$, $[12] = \epsilon^{\dotA\dotB} \bu_{1\dotA}
\bu_{2\dotB}$. 
The scattering amplitudes are expressed as functions of the 
invariant scalar products of spinor-momenta.

The amplitude for the scattering of $n-2$ gluons of negative
helicity and $2$ gluons of positive helicity is given by
\beqar
{\cal A} (1^{a_1}_{-}, 2^{a_2}_{-}, 3^{a_3}_{+}\cdots n^{a_n}_{+})
&=&i g^{n-2}  ~
\delta^{(4)}\left(\sum_i p_i\right)~\Tr (t^{a_1} t^{a_2}\cdots t^{a_n})~{\cal M}\nonumber\\
{\cal M}(1,2,...,n)&=& {\la 12\ra^4  \over
\la 12\ra \la 23\ra \cdots \la n-1~n\ra \la n 1\ra}
\label{3}
\eeqar
where $g$ is the coupling constant and 
$1,2$ refer to the negative helicity gluons.
For simplicity of presentation, all gluons are taken as incoming.
The expression ${\cal A}$ in (\ref{3})
is actually a subamplitude, the full amplitude is obtained by
summing over such subamplitudes with all noncyclic permutations.
(This subamplitude has cyclic symmetry, 
so we can also sum over all permutations and divide by $n$.)

The spin operator is given by $S_\mu \sim \epsilon_{\mu\nu\alpha\beta}
J^{\nu\alpha}p^\beta$, where $J^{\mu\nu}$ is the Lorentz generator.
This can be simplified for null momenta and the helicity is then
identified as
\beq
s = - {1\over 2} u^A {\del \over \del u^A}
\label{5}
\eeq
Thus $-s$ is half the degree of homogeneity in the $u$'s.
If we start with a positive helicity gluon, then we should expect
an additional four factors of $u$ for a negative helicity gluon.
This is in keeping with the factor $\la 12\ra^4$ in (\ref{3}).

A useful way of simplifying this is to notice that,
if we have an anticommuting spinor
$\theta^A$, $\int d^2 \theta ~\theta_A \theta_B = \epsilon_{AB}$,
so that
\beq
\int d^8\theta ~\prod_{\alpha =1}^4 \theta^\alpha_A u_1^A~\prod_{\beta =1}^4
\theta^\beta_B u_2^B = \la 12\ra^4\label{6}
\eeq
We then see that the factor $\la 12\ra^4$ can be naturally
represented
in an ${\cal N}=4$ Yang-Mills theory.
In fact, analogous to the spinor momenta, we can associate an anticommuting
spinor variable $\eta^\alpha_i$ for the $i$-th particle, where
$\alpha =1, ...,4$. We can introduce a Dirac delta function
for these variables and combine the momentum conservation and the 
factor of $\la 12\ra^4$ as
\beq
\delta^{(4)} \left(\sum_i p_i\right) \delta^{(8)} \left(\sum u_i\eta_i \right)
= \int d^4x d^8\theta~ \prod_i \exp \left( i \bu_{\dot A i} x^{\dot A}_A u^A_i
+ i \theta^\alpha_A u^A_i \eta^\alpha_i \right)
\label{7}
\eeq 
The factor $\exp \left( i \bu_{\dot A i} x^{\dot A}_{~A} u^A_i
+ i \theta^\alpha_A u^A_i \eta^\alpha_i \right)$ may be considered as part of the wave function
of the $i$-th particle, which corresponds to the full ${\cal N}=4$ supermultiplet.
The powers of $\eta$ define the different helicity components in accordance with
(\ref{5}), since an
expansion in the $\eta$'s generate powers of $u$.
For two gluons of negative helicities and $n-2$ gluons of positive helicities,
we recover the factor $\la 12\ra^4$. Other powers of $\eta$ will correspond to 
the MHV amplitudes for the other particles in the ${\cal N}=4$ supermultiplet.

The denominator in (\ref{3}) can be obtained in terms of current correlators constructed from
free fermions.
We can introduce a current ${\cal J} (1) = \alpha (1) \beta (1)$, where
$\alpha$, $\beta$ are two fermion fields defined on the 
space of the spinor momenta by
\beq
\la \beta (1)~ \alpha (2) \ra = {1\over \la 12\ra}
\label{8}
\eeq
As is well known, these are essentially fermion fields
on 
the complex projective space ${\bf CP}^1$, corresponding to the
spinor momenta, up to a complex scale transformation.
This space is defined by $u^A$, with
$u \sim \lambda u $,
for any complex number $\lambda$ which is not zero,
$\lambda \in {\bf C} - \{ 0\} $.
For a local coordinate parametrization, if we write,
\beq
u = \left( \matrix{ a\cr b\cr}\right)
\label{10}
\eeq
we can take
$b/a =z$ as the local complex coordinate
of ${\bf CP}^1$ except in the neighborhood of
$a =0$; near $a =0$, we can use $a/b$
as the local coordinate.
One can also define a local version of the  fields by $\psi = a \alpha$ and
$\psi ' = a \beta$. 
The amplitude (\ref{3}) can be obtained in terms
of the current correlators
$\la {\cal J} (1) {\cal J} (2) \cdots {\cal J} (n)\ra$.
For the purpose of comparison with graviton amplitudes, it is easier to express
the amplitude in terms of $\bdel^{-1}$, the inverse of the $\bdel$-operator
on the ${\bf CP}^1$ of spinor momenta.
This is given by
\beq
G(1, 2) = G_{12}= \left({1\over \bdel}\right)_{12} = - {1\over \pi} {1\over \la 12\ra}
\label{28}
\eeq
(On functions with degree of homogeneity
equal to $1$,  the right hand side simplifies to
$1/ \pi (z_1 - z_2)$ in local coordinates.)
We also introduce the gauged version $\bD = \bdel - \bA$ where
\beqar
\bA_i&=&g\pi  t^a \vf^a_i ~\exp \left( i \bu_{\dot A i} x^{\dot A}_A u^A_i
+ i \theta^\alpha_A u^A_i \eta^\alpha_i \right)\nonumber\\
&\equiv&g\pi  t^a \vf^a_i ~V_i
\label{12}
\eeqar
Using these, we can finally write the MHV amplitudes for ${\cal N} =4$ Yang-Mills theory as
\beqar
{\cal A} (1^{a_1}, 2^{a_2}, 3^{a_3}, \cdots ,n^{a_n})
\!\!\!&=&\!\!\! i \left[{\delta \over \delta \vf^{a_2}_2}{\delta \over \delta \vf^{a_3}_3}
\cdots {\delta \over \delta \vf^{a_n}_n}~W[\vf ,1] \right]_{\vf =0}\nonumber\\
\!\!\!&=&\!\!\! i g^{n-2} (-\pi )^n ~
\Tr (t^{a_1} t^{a_2}\cdots t^{a_n}){\cal M}(1,2,...,n)\nonumber\\
&&\hskip .3in~+{\cal P}(2,3,...,n)
\label{13}
\eeqar
where ${\cal P}(2,3,...,n)$ denotes permutations of the labels $2,3,...,n$ and 
\beqar
W[\vf ,1] &=& {1\over g^2} \int d^4x d^8\theta ~ \Tr\left[ \bA_1\left( {1\over \bD}\right)_{11}\right]_{\vf^{a_1}_1=1}\nonumber\\
&=& -{1\over g^2}{\delta \over \delta \vf^{a_1}_1} \int d^4x d^8\theta ~\Tr \log (\bdel -\bA )\label{13a}\\
{\cal M}(1,2,...,n)&=&
\int d^4xd^8\theta~
G_{12} V_2 G_{23}\cdots G_{n1}
\label{13b}
\eeqar
Finally, we note that from the point of view of holomorphicity in
twistor space,
a key step 
is the Fourier transform introduced in \cite{witten},
\beqar
{\cal M} &=& \int \prod_i d^2\omega_i d^4\psi_i  
\exp ( i \omega^{\dot A}_i \bu_{\dot A i} +i \psi^\alpha _i \eta^\alpha_i )~
\int d^4xd^8\theta~ {\widetilde{\cal  M}}\nonumber\\
{\widetilde{\cal M}}&=& \prod_i \delta ( \omega^{\dot A}_i - x^{\dot A}_A u^A_i )
\delta ( \psi^\alpha_i - \theta^\alpha_A u^A_i )~ 
G_{12} V_2 G_{23}\cdots G_{n1}
\label{14}
\eeqar
${\widetilde{\cal M}}$ is holomorphic in all the twistor variables
$Z^\alpha= (\omega^{\dot A}, u^A), ~\psi^\alpha$, and has support, by virtue of the
$\delta$-functions,
on the line $\omega^{\dot A} - x^{\dot A}_{~A} u^A =0$, $ \psi^\alpha - \theta^\alpha_A u^A=0$.
In constructing the scattering amplitude ${\cal M}$,
we must integrate over the moduli $x^{\dot A}_A$ and $\theta^\alpha_A$ of this line,
in addition to the Fourier transformation.
(The Fourier transformation is clear if we use the 
$(++--)$ signature; in other cases, the Fourier transformation has to be interpreted
more carefully \cite{witten}.)

The generalization to non-MHV amplitudes given in \cite{witten}
is to make a similar construction
by considering curves of higher  degree in twistor space;
as mentioned before, this has been verified 
by a number of explicit calculations \cite{list1}.

We shall now turn to the question: Can we do a similar simplification and 
vertex operator representation for the MHV amplitudes for gravitons?
The MHV amplitude for two negative helicity gravitons and $n-2$ positive helicity 
gravitons is given by \cite{berends}
\beqar
{\cal A} (1_{-}, 2_{-}, 3_{+}, \cdots , n_{+})
&=& \left( {\kappa \over 2}\right)^{n-2} \delta^{(4)}\left(\sum_i p_i\right)~ {\cal M}
(1_{-}, 2_{-}, 3_{+}, \cdots , n_{+})\nonumber\\
{\cal M} (1_{-}, 2_{-}, 3_{+}, \cdots , n_{+})&=& 
\la 12\ra^8 \Biggl[ {[12] [n-2 ~n-1] \over \la 1~n-1\ra} {1\over N(n)} \prod_{i=1}^{n-3}
\prod_{j=i+2}^{n-1} \la ij\ra ~F \nonumber\\
&&\hskip .5in+{\cal P} (2,3,...,n-2)\Biggr]
\label{15}
\eeqar
where $N(n) = \prod_{i,j, i<j} \la ij\ra$ and 
\beq
F= \left\{ \begin{array}{ll}
\prod_{l=3}^{n-3} \bu_{\dot A l} (p_{l+1} + p_{l+2}+\cdots +p_{n-1})^{\dot A} _A u^A_n~~~~&
n\geq 6\\
1& n=5\\
\end{array}
\right.\label{16}
\eeq
As before, ${\cal P}(2,3,..., n-2)$ indicates permutations of the labels $2,3,..., n-2$, 
and $\kappa$ is given in terms of Newton's constant $G$ as
$\kappa = \sqrt{32\pi G}$.
The formula given above is for $n\geq 5$; for $n=4$, Berends {\it et al} give a separate
formula
\beq
{\cal M} (1_{-}, 2_{-}, 3_{+}, 4_{+})
= {\la 12\ra^8 [12] \over N(4) \la 34\ra }
\label{17}
\eeq
These formulae are algebraically somewhat involved; they also have many factors
which are not holomorphic in the spinor momenta. Their representation in 
twistor language will therefore be more complicated than in the Yang-Mills case.
In the Fourier-transformed version, as in (\ref{14}), one can represent
$\bu_{\dot A}$ as a derivative with respect to $\omega^{\dot A}$ and so, one can have a holomorphic function in twistor space, but the localization to the curve will now involve derivatives of the delta function
$\delta ( \omega^{\dot A}_i - x^{\dot A}_A u^A_i )$.
Rather than follow such a procedure, we shall first simplify the
expressions (\ref{16}), (\ref{17}).

First we shall consider the four-point function. The factor  $N(4)$
can be written
as $ - C(4) \la 13\ra \la 24\ra$, where $C(4)$ is the cyclically invariant
combination $\la 12\ra \la 23\ra \la 34\ra \la 41\ra$. Next we notice that,
by virtue of momentum conservation,
\beqar
{[2 P4\ra \over \la 24\ra}&=&{\bu_{\dot A 2} P^{\dot A}_A u^A_4 \over \la 24\ra}\nonumber\\
&=& - {\bu_{\dot A 2} (p_1)^{\dot A}_A u^A_4 \over \la 24\ra}\nonumber\\
&=& - {[12] \la 41\ra \over \la 24\ra}
\label{18}
\eeqar
In this equation, $P$ stands for $p_2 + p_3 +p_4$, which is the total momentum of all
particles to the right of the particle labeled $2$ in the chosen ordering.
Using (\ref{18}) and the expression for $N(4)$ we can write
\beq
{\cal M} (1_{-}, 2_{-}, 3_{+}, 4_{+})
= \la 12\ra^8 ~ \left[{1\over \la 12\ra \la 23\ra \la 34\ra \la 41\ra}\right]
\left[{[2 P4\ra \over \la 24\ra} {1\over \la 13\ra \la 34\ra \la 41\ra}\right]
\label{19}
\eeq
As in the case of the Yang-Mills fields, we may expect the first factor
$\la 12\ra^8$ to be obtained via integration over anticommuting parameters.
The second factor, with the cyclically symmetric denominator, can be represented
in terms of current correlators, or $\bD^{-1}$, as in the Yang-Mills case.
The last factor involves an expression similar to the current correlator, but
with only three particles, $1, 3, 4$ participating.
To write this in a compact way, we
introduce another set of free fermions, $\chi$ and $\phi$, with
\beq
\la \phi (1) \chi (2) \ra = \la \phi_1 \chi_2 \ra = {1\over \la 12\ra}
\label{20}
\eeq
We may think of these fields as having the mode expansion
\beq
\alpha \chi (z) = \sum_{n=0} a^\dagger_n z^{-n-1}, \hskip .5in
\alpha \phi (z) = \sum_{n=0} a_n z^n \label{21}
\eeq
in the local coordinate $z = b/a$. $a^\dagger_n, a_n$ are creation and 
annihilation operators and we may assign a vacuum for 
these fermions by $a_n \vert 0\ra =0$.
In the last factor in the
expression (\ref{19}), there is no independent symmetrization over
$1, 3, 4$. We can obtain the last two factors in
(\ref{19}), with the specific order chosen, by writing
\beq
{[2 P4\ra \over \la 24\ra} {1\over \la 13\ra \la 34\ra \la 41\ra}
= \int {d^2z_\lambda\over \pi} ~ \la 0\vert \phi_\lambda
(\chi \phi )_1 {[2 P\lambda\ra \over \la 2\lambda\ra} (\chi \phi )_3
(\chi \phi )_4 \bdel_\lambda \chi_\lambda \vert 0\ra
\label{22}
\eeq
We get a delta function by the action of $\bdel_\lambda$ on the correlator
$\la \phi_4 \chi_\lambda \ra$ which sets the point $\lambda$ to the
spinor momentum of particle $4$ upon integration over $z_\lambda$.
An alternative way to represent this is via a Penrose contour integral.
If $f(\lambda )$ has zero degree of homogeneity in $\lambda$,
\beq
\oint_{C_4}  {\epsilon_{AB} \lambda^A d \lambda^B \over 2\pi i}
{1\over \la 4\lambda\ra \la \lambda 1\ra } f (\lambda ) 
= {1\over \la 41\ra} f(4) 
\label{23}
\eeq
where the contour encloses the pole at $z_\lambda = z_4$.
This result can be verified by direct integration.
Using (\ref{23}),
\beqar
{[2 P4\ra \over \la 24\ra} {1\over \la 13\ra \la 34\ra \la 41\ra}
&=& \oint_{C_4} ~ \la 0\vert \phi_\lambda
(\chi \phi )_1 {[2 P\lambda\ra \over \la 2\lambda\ra} (\chi \phi )_3
(\chi \phi )_4\chi_\lambda \vert 0\ra\nonumber\\
&=&\oint_{C_4} ~ \la \lambda\vert
(\chi \phi )_1 {[2 P\lambda\ra \over \la 2\lambda\ra} (\chi \phi )_3
(\chi \phi )_4 \vert \lambda\ra
\label{24}
\eeqar
It is understood that the integration is done with the measure
${\epsilon_{AB} \lambda^A d \lambda^B / 2\pi i}$. For brevity, we also
use $\vert \lambda \ra = \chi_\lambda \vert 0\ra$.

The four-graviton MHV amplitude may now be written as
\beqar
{\cal A} (1_{-}, 2_{-}, 3_{+}, 4_{+})
&=& - \left( {\kappa \over 2}\right)^{2}   {\la 12\ra^8\over \la 12\ra \la 23\ra \la 34\ra \la 41\ra}
\nonumber\\
&&\times \int d^4x \oint_{C_4}  \la \lambda\vert 
{\widetilde V}_1 {\widetilde {\cal E}}_2 {\widetilde V}_3 {\widetilde V}_4\vert\lambda\ra
\label{25}
\eeqar
where
\beqar
{\widetilde V} &=&\chi \phi ~\exp ({ip\cdot x})\nonumber\\
{\widetilde{\cal E}} &=& \exp ({ip\cdot x})~ {\bu_{\dot A} \lambda^A (-i \nabla^{\dot A}_{~A}) \over
\la u\lambda\ra }\nonumber\\
&=&{\bu_{\dot A} \lambda^A (-i \nabla^{\dot A}_{~A}) \over
\la u\lambda \ra}~\exp ({ip\cdot x})
\label{26}
\eeqar
where $\la u\lambda \ra = \epsilon_{CD}u^C\lambda^D$.
The differentiation of the factors $e^{ip\cdot x}$ to the right of ${\widetilde {\cal E}}$
will produce the factor $P = (p_2 +p_3 +p_4)$ in the formula (\ref{19}).

We have carried out the simplification of the four-graviton amplitude in some detail.
For amplitudes involving higher number of gravitons, a similar simplification can  be done
and it can be verified that they can all be represented as
\beqar
{\cal A} (1_{-}, 2_{-}, 3_{+}, \cdots , n_{+})
&=&-  \left( {\kappa \over 2}\right)^{n-2} \la 12\ra^8
\int d^4x \Biggl[ { 1\over \la 12\ra \la 23\ra \cdots \la n1\ra} \times\nonumber\\
&& \oint_{C_n} 
\la \lambda \vert {\widetilde V}_1 {\widetilde {\cal E}}_2 {\widetilde {\cal E}}_3 \cdots
{\widetilde {\cal E}}_{n-2} {\widetilde V}_{n-1} {\widetilde V}_n\vert\lambda\ra\nonumber\\
&&\hskip .4in +~{\cal P} (2,3,..., n-2)\Biggr]
\label{27}
\eeqar
Expressing the factor $1/C(n)$ in terms of $\bdel^{-1}$, we can further simplify this as
\beq
{\cal A} (1_{-}, 2_{-}, 3_{+}, \cdots , n_{+})
=- {4\over \kappa^2} \left( -{\kappa \pi \over 2}\right)^n~ \la 12\ra^8~ {\cal F}(1, 2, \cdots , n)\label{27a}
\eeq
where
\beqar
{\cal F}&=& \int d^4x
\oint_{C_n} \Biggl[
\la \lambda \vert {\widetilde V}_1 G_{12} {\widetilde {\cal E}}_2
G_{23} \cdots
{\widetilde {\cal E}}_{n-2} G_{n-2~n-1}{\widetilde V}_{n-1}
G_{n-1 ~n}{\widetilde V}_n G_{n1}\vert\lambda\ra\nonumber\\
&&\hskip 1.5in~+~{\cal P} (2,3,..., n-2)\Biggr]
\label{29}
\eeqar
We can go one more step in the reduction of this expression by introducing 
the supersymmetric vertex operators
\beqar
{V} &=&- {\kappa \pi \over 2}~v~\chi \phi ~\exp ({ip\cdot x} + {iu^A \theta^\alpha_A\eta^\alpha}) \nonumber\\
{\cal E}&=& -{\kappa\pi\over 2} ~h~\exp ({ip\cdot x} + {iu^A \theta^\alpha_A\eta^\alpha})~ {\bu_{\dot A} \lambda^A (-i \nabla^{\dot A}_{~A}) \over
\la u\lambda \ra}\label{30}\\
\bA&=& V+ {\cal E}\nonumber
\eeqar
where $\alpha =1, 2, ..., 8$ in the present case of gravitons.
We are thus considering supergraviton amplitudes in ${\cal N}=8$
supergravity.
The factor of $\la 12\ra^8$ is obtained by integration over the $\theta$'s if we pick up the
term corresponding to the first two gravitons having negative helicity, which is
the term with $8$ powers of $\eta$ for each, or equivalently $8$ powers of
$u_1$ and $u_2$.
Combining everything, we can write the MHV amplitude for
graviton supermultiplets as
\beqar
{\cal A}(1, 2, \cdots , n)\!\!\! &=&\!\!\!\Biggl[{1\over 2!}
{\delta \over \delta h_2}\cdots
{\delta \over \delta h_{n-2}}{\delta \over \delta v_{n-1}}{\delta \over \delta v_n}~W[h,v,1]\Biggr]_{\bA=0}~\label{31}\\
W[h,v,1]\!\!\!&=&\!\!\! -{4\over \kappa^2}\int d^4xd^{16} \theta\Biggl[\oint_{C_n} 
\la \lambda \vert  V_1 \left({1\over \bdel - \bA }\right)_{11}  \vert\lambda\ra \Biggr]_{v_1 =1}
\label{32}
\eeqar
The functional derivatives in (\ref{31}) give symmetrization with respect to 
the labels $2,3,..., {n-2}$. There is also symmetrization with respect to
$n-1$ and $n$, which is redundant, since the expression is symmetric
in all labels once the symmetrization with respect to $2, 3, ..., n-2$ is carried out;
the factor $1/2!$ avoids double counting due to this.
In the expansion of the expression for $W[h,v,1]$ in powers of $V$ and ${\cal E}$,
notice that we need $V$'s for the last two places, namely for $n-1$ and $n$.
Otherwise the expression gives zero, since ${\cal E}_{n-1}\sim  [n-1 (p_{n-1}+p_n)n\ra
=0$, ${\cal E}_n \sim [n p_n n\ra =0$.
Equations (\ref{31}), (\ref{32}) are thus seen to lead back to (\ref{27a}), (\ref{29}).
The formula picks out three gravitons as different from the rest.
In the end, it does not matter which three, since the symmetrization
over the labels $2, 3,..., n-1$ will automatically make the expression
in, say (\ref{27}), totally symmetric, apart from the factor $\la 12\ra^8$.
In (\ref{32}), we are picking out three vertices of the $V$-type and $n-3$
of the ${\cal E}$-type. In principle, one could consider amplitudes
obtained by having more vertices of the $V$-type, equivalently
having more derivatives with respect to $v$'s. It is not clear at this 
stage what the physical interpretation of such amplitudes would be.

We now turn to an interpretation of the vertex operators $V$, ${\cal E}$, or
$\bA = V+{\cal E}$. In the Yang-Mills case, the vertex operator $\bA$ given in
(\ref{12}) was the helicity-one projection of the gauge potential $A^A_{~\dot A}$,
which we may write as 
$\bu^{\dot A} (D^A_{~\dot A} - \nabla^A_{~\dot A})/\la u\lambda\ra$.
In the gravitational case we may therefore expect the helicity-one projection
of the covariant derivative with gravitational fields.
This is given by
\beq
D^A_{~\dot A} = (\sigma^\mu)^A_{~\dot A} \left[ e^a_\mu \del_a - \omega^{ab}_{\mu}
J^{ab} \right]
\label{33}
\eeq
where $e^a_\mu, ~\omega^{ab}_\mu$ are the tetrad field and the spin connection,
$J^{ab}$ being the Lorentz generator. For perturbative graviton calculations, we may write
$e^a_\mu \approx \delta^a_\mu - h^a_\mu$, and
\beq
D^A_{~\dot A} \approx  \nabla^A_{~\dot A}
- \left( h^a \del_a + \omega^{ab}
J^{ab} \right)^A_{~\dot A}
\label{34}
\eeq
The helicity-one projection of $(h^a \del_a )^A_{~\dot A}$ gives a term like
${\cal E} $ in (\ref{30}). In the spin connection term, by choosing a gauge
where $h^a_\mu$ has zero divergence and zero trace, which is appropriate for
on-shell gravitons, one can see that the helicity-one projection
does not give any additional $h$-dependent terms.
However, in supergravity, the spin connection has terms which are quadratic
in the gravitino fields; the helicity-one projection of this term leads to
an expression of the form $\chi \phi$, where $\chi$, $\phi$ are appropriate
components of the gravitino. This could account for the vertex operator
$V$ in (\ref{30}).
$\bA = V +{\cal E}$ has values in the Poincar\'e algebra and, in analogy with
the gauge field case, and from what has been said above, we see that the
$V, ~{\cal E}$ terms can be considered as the Chan-Paton factors.
The full vertex operator for gravitons is then of the form
$J~ V$ or $J~ {\cal E}$, where $J$ is the current constructed
as $\alpha \beta$, identical to Yang-Mills case and leading to the
$\bdel^{-1}$ factors. The contribution of the
$J$'s is naturally given in twistor language.

We may note that
some of the amplitude calculations done in Berkovits' string theory
have a similar structure, although that theory has conformal supergravity, rather than
Einstein supergravity \cite{witt-berk}.
It is possible that a variant of this theory can lead to the vertex operators
we have discussed.
\vskip .1in\noindent

The results in this note were presented at the London Mathematical Society 
Workshop on Twistor String Theory, University of Oxford, January 10-14, 2005.
This work
was supported by the National Science Foundation grant number
PHY-0244873 and by a PSC-CUNY award. I thank L. Dixon for bringing
reference \cite{seliv} to my attention.

\end{document}